\documentclass[twocolumn,showpacs,preprintnumbers,amsmath,amssymb %,superscriptaddress
 amsmath,amssymb,
 prb,
 lengthcheck,%
]{revtex4-1}

\usepackage{graphicx}% Include figure files
\usepackage{dcolumn}% Align table columns on decimal point
\usepackage{bm}% bold math
\usepackage{hyperref}% add hypertext capabilities

\usepackage{makeidx}
\makeindex
\def\>{\right\rangle}
\def\<{\left\langle}
\def\be{\begin{equation}}
\def\ee{\end{equation}}
\def\ba{\begin{array}{l}}
\def\ea{\end{array}}
\def\f{\frac}

\def\beq{\begin{eqnarray}}
\def\eeq{\end{eqnarray}}

\begin{document}

\preprint{APS/123-QED}

\title{Charge tunneling in fractional edge channels}

\author{D. Ferraro$^{1}$, A. Braggio$^{2}$, N. Magnoli$^{1}$, M. Sassetti$^{3}$}
 \affiliation{$^1$ Dipartimento di Fisica e INFN, Universit\`a di Genova,Via Dodecaneso 33, 16146, Genova, Italy.\\
$^2$  CNR-SPIN, Dipartimento di Fisica Via Dodecaneso 33, 16146, Genova, Italy\\ 
$^3$ Dipartimento di Fisica e CNR-SPIN,  Via Dodecaneso 33, 16146, Genova, Italy.}
\date{\today}% It is always \today, today,

\begin{abstract}
  We explain recent experimental observations on effective charge of
  edge states tunneling through a quantum point contact in the weak
  backscattering regime.  We focus on the behavior of the excess
  noise and on the effective tunneling charge as a function of temperature
  and voltage.  By introducing a minimal hierarchical model
  different filling factors, $\nu=p/(2p+1)$, in the Jain sequence are
  treated on equal footing, in presence also of non-universal interactions. The agreement found with the experiments for
  $\nu=2/3$ and $\nu=2/5$ reinforces the description of tunneling of
  bunching of quasiparticles at low energies and quantitatively
  defines the condition under which one expects to measure the
  fundamental quasiparticle charge.
  We propose high-order current cumulant measurement to
  cross-check the validity of the above scenario and to better
  clarify the peculiar temperature behavior of the effective charges measured in the experiments.
\end{abstract}

\maketitle

\section{Introduction}

Fractional quantum Hall effect represents one of the most important
examples of strongly correlated electron system \cite{DasSarma97}. In
the bulk, quasiparticle (qp) excitations are predicted to have
 fractional charge \cite{Laughlin83} which, e.g., for
filling factor in the Jain series, $\nu = p/(2p + 1)$ ($p\in
\mathbb{Z}$), is $e^{*} = e(\nu/|p|)$.  At the edge \cite{Wen90,Wen91,Wen95}
the identification of these charge excitations seems more
complicated. Indeed, while in the past measurements of current noise through
quantum point contacts (QPC), in the weak 
backscattering regime, confirmed the
tunneling of single-qp\cite{dePicciotto97, Seminadayar97}, recently,
new measurements have demonstrated
the possibility
of tunneling charges multiple of the fundamental charge. The
condition to observe a bunching of qp depends on the external
parameters such as temperature and voltage. Measurements\cite{Chung03} carried out
for the Jain series ($p=2,3$), at extremely low temperatures,
show an effective charge equal to $e_{\rm eff}=
\nu e$, which, only by increasing the temperature, decreases to the fundamental value $e_{\rm eff}=e^{*}$.  Last
year, experimental results for filling factor $\nu=2/3$ $(p=-2)$ 
appeared \cite{Ofek09}, showing a similar crossover. This common trend was
very recently verified also for filling factor outside the Jain
series belonging to fractional values in the second Landau level
\cite{dolev10}.  

In addition to the bunching phenomena peculiar
behavior also appears in the backscattering current at high transparencies. For example for $\nu=1/3$, the current was found to increase with temperature
\cite{Chung03,Roddaro04} instead of decrease as theoretically predicted \cite{Fendley95}. This
support the indication of a non-universal renormalization of the
tunneling exponents induced by the presence of edge interaction with
external environment \cite{Rosenow02},
electron-electron interaction \cite{Papa04, Mandal02} and edge
reconstruction \cite{Yang03,Aleiner94}.  

In order to describe the Jain sequence different models were proposed
with the common requirement of the presence of neutral modes in order
to fulfill the statistical properties. One could have $|p|-1$ neutral fields 
propagating at finite velocity along the edge \cite{Wen92,Kane94,Kane95}, 
or only two or one - for infinite edges - additional modes with zero \cite{Lopez99, Lopez01} or
finite velocity \cite{Ferraro08}.  A peculiar characteristic, associated to the
neutral modes is their direction of propagation with respect to
the charged mode.  Depending on the sign of $p$ and the theoretical model, there is 
the possibility to have co-propagating or counter-propagating neutral modes.

The tendency of bunching of qp at low temperature and weak
backscattering was underlined in theory for the hierarchy of the Jain
sequence \cite{Kane94,Kane95,Ferraro08}.  In Ref.~\onlinecite{Ferraro08}
we pointed out the role of propagating neutral modes in order
to fully describe the experimental data \cite{Chung03} for $p>1$.
%, showing tunneling between a bunching of qp at low energies and a single-qp at higher
%energies. 
By comparing with experiments for $\nu=2/5$ it was indeed possible
to estimate the energy bandwidth of neutral modes.  

Despite the presence of different proposals on the direct detection of neutral
modes \cite{Levin07,Feldman08,Ferraro08,Overbosch09,Ferraro09b,Cappelli09,Cappelli10,Yang09}, experiments addressed this issue only recently \cite{Granger09,Aveek10}.

In this paper we present a minimal hierarchical model able to
include all the essential features of the above different proposal
using few free parameters. This allows to explain,
in an unified background, the experimental results of tunneling of
effective charges in a standard quantum point contact geometry at extremely high
transmission \cite{Chung03, Ofek09}.  The dependence of the
excess noise on the external parameters such as the voltage and the
temperature is quantitatively analyzed. The flexibility
of the proposed model resides on the possibility to
link the results obtained in the presence of counter-propagating or
co-propagating neutral modes. We demonstrate that
both cases reproduce the
experimental results using a proper choice of the fitting parameters.

We also propose the skewness, namely the normalized third
backscattering current cumulant, as a measurable quantity
\cite{Reulet03,Lindell04,Bomze05,Huard07,Timofeev07,Gershon08} able to
give independent information on the nature of the carriers. This quantity is a 
good estimator of the crossover in the tunneling between the bunching of qp and  the
 fundamental charge. We show that this quantity can be directly compared 
 with the {\textit{effective charge}} measured in the experiments by fitting the excess noise, as a function of the bias voltage, at fixed values 
of temperature.

\section{Model}
We consider infinite edge states of an Hall bar with  filling factor in the Jain series 
$\nu=p/(2p+1)$ ($p\in \mathbb{Z}$). The model adopted is  a minimal one with two decoupled bosonic fields, one charged
$\varphi^{\rm{c}}$ and one neutral $\varphi^{\rm{n}}$. The Euclidean
free action is ($\hbar=1$, $k_{\rm B}=1$)
\beq
\mathcal{S}^{0}&&=\f{1}{4\pi\nu} \int^{\beta}_{0} \!\!\!d\tau \!\!\int^{+\infty}_{-\infty} \!\!\!\!\!\!dx \partial_{x}\varphi^{\rm{c}}(x,\tau)\left(i
\partial_{\tau}+v_{\rm{c}}\partial_{x}\right)\varphi^{\rm{c}}(x,\tau)+
\nonumber\\
&&+\f{1}{4\pi} \int^{\beta}_{0} \!\!\! d\tau \!\!\!\int^{+\infty}_{-\infty}\! \!\!\!\!\!\!dx \partial_{x}\varphi^{\rm{n}}(x,\tau)\left(i\xi \partial_{\tau}+v_{\rm{n}}
\partial_{x}\right)\varphi^{\rm{n}}(x,\tau)\,,
\label{action}
\eeq 
with $\beta=T^{-1}$ the inverse temperature and $v_{\rm{c}}$,
$v_{\rm{n}}$ the propagation velocities of charge and neutral
modes respectively. The former is affected by Coulomb interactions
\cite{Levkinskyi08, Levkinskyi09} such that $v_{\rm{c}}\gg v_{\rm{n}}$
\cite{Ferraro08}. We consider neutral modes co-propagating $(\xi=+1)$
or counter-propagating $(\xi=-1)$ with respect to the charged one.
This choice allows a unified description of different hierarchical models. For $\xi={\rm{sgn}}(p)$ one recovers the
restricted model of Lee and Wen \cite{Lee98} (LW), where the $|p|-1$
neutral modes are described in terms of a single one. While for
$\xi=-{\rm{sgn}}(p)$ one obtains the generalized Fradkin-Lopez model
\cite{Lopez99, Chamon07, Ferraro08, Ferraro09b} (GFL) with a single
neutral mode propagating at finite velocity instead of a topological one \cite{Lopez99}.

The commutators of  the bosonic fields are $[\varphi^{\rm{c/n}}(x),\varphi^{\rm{c/n}}(y)]=i\pi \nu_{\rm{c/n}} 
\mathrm{sgn}(x-y)$ with $\nu_{\rm{c}}=\nu$ and $\nu_{\rm{n}}=\xi$. The electron number density depends on the charged 
field only, via the relation $\rho(x)=\partial_{x}\varphi^{\rm{c}}(x)/2\pi$.

\textit{Edge excitations.} 
In the hierarchical theories admissible edge excitations have a well defined charge and statistics \cite{Wen92, Lopez99}. There are single-qp excitations with charge 
 $e^{*}$ with $ e^{*}=(\nu/|p|)e$ and multiple qp-excitations with charge $m e^{*}$ ($m\in \mathbb{N}$)\cite{Note1}.  Their statistics is fractional with statistical angle \cite{Su86}
\be
\theta_{m}=m^{2}\left( \f{\nu}{p^{2}}-\f{1}{p}-1\right) \pi\,\,\,\,\,\,\, (\rm{mod} \,\,2 \pi).
\label{stat}
\ee
In addition, the phase acquired by any excitation in a loop around an electron must be an integer multiple of $2\pi$ \cite{Froehlich97, Ino98, Ferraro09b}.
Using the bosonization technique and imposing the above constraints,
one can write the $m$-multiple excitation operator \cite{Ferraro09b}
\be
\Psi^{(m,q)}(x)=\f{\mathcal{F}^{(m,q)}}{\sqrt{2\pi a}}e^{i\left\{\left(s+\f{d}{|p|}\right)\varphi^{\rm{c}}(x)+\sqrt{p^{2}-\xi p}
\left(q+\f{d}{|p|}\right)\varphi^{\rm{n}}(x)\right\}}
\label{qp_operator}
\ee with $a$ cut-off length, $s\in \mathbb{N}$ and $0\leq d\leq |p|-1$
such that $m=s|p|+d$. The integer $q$ is an additional quantum number associated to the freedom of add $2\pi$ to the statistical
angle \cite{Ferraro09b}.  The operator
$\mathcal{F}^{(m,q)}$ changes the number of $m$-agglomerates on the
edge and ensures the right statistical properties between different
$q$-values and different edges\cite{Ferraro09b} . It can be neglected in the
sequential tunneling regime \cite{Ferraro09b, Guyon02, Martin05}.  The
most general expression for an excitation with charge $me^*$ will be
then given by a superposition of the above operator with different $q$
values \cite{Ferraro09b, Wen95}.

\textit{Relevant excitations.} The scaling dimension associated to an
$(m,q)$-excitation is extracted from the long time limit of the
two-point imaginary time Green's function $\mathcal{G}^{(m,q)}(
\tau)=\<T_{\tau}\Psi^{(m, q)} (0,\tau){\Psi^{(m,q)}}^{\dagger}(0,0)\>$
at zero temperature \cite{Kane92}.  For $ |\tau | \gg
\omega_{\rm{n}}^{-1}, \omega_{\rm{c}}^{-1}$ it is $
\mathcal{G}^{(m,q)}( \tau)\propto |\tau|^{-2 \Delta_{m}(q)}$ with \be
\Delta_{m}(q)=
\frac{g_{\rm{c}}\nu}{2}\!\left(\f{m}{|p|}\right)^{2}\!\!\!+ \!
\frac{g_{\rm{n}}}{2} (p^{2}-\xi p)\! \left(q+ \f{d}{|p|}\right)^{2}.
\label{scaling}
\ee Here, $\omega_{\rm{c,n}}=v_{\rm{c,n}}/a$ are the energy bandwidth
and satisfy $\omega_{\rm{n}}\ll \omega_{\rm{c}}$. The first term in (\ref{scaling}) is due to the charged mode, while the second is related to the neutral one. The parameters $g_{\rm
  c}$ and $g_{\rm n}$ are introduced to take into account possible
interaction effects due to the external environment \cite{Rosenow02,
  Papa04, Mandal02, Yang03}. It is worth to note that the two models
considered, with $\xi=\pm {\rm sgn}( p)$, differ in the neutral mode contribution
only. However, introducing neutral renormalization parameters $g^{\rm{
    LW}}_{\rm{n}}$ and $g^{\rm{GFL}}_{\rm{n}}$ for the LW and the
GFL model respectively, one can map the two cases via the substitution
\be
\label{eq:mapping}
g^{\rm{LW}}_{\rm{n}}=g^{\rm{GFL}}_{\rm n}\ \frac{p^{2}+|p|}{p^{2}-|p|}\ .
\ee
Operators with the minimal scaling dimension are the most relevant and dominate the transport properties at low energies $E\ll \omega_{\rm n},\omega_{\rm c}$ \cite{Kane92, Ferraro08, Ferraro09b}. In the unrenormalized case $(g_{\rm c}=g_{\rm n}=1)$ the two most 
dominant excitations have always $q=0$. They correspond to the agglomerate with $m=|p|$ ($d=0$, $s= 1$)  
and to the single-qp  with $m= 1$ ($d= 1$, $s=0$). The corresponding scaling are
\be 
\Delta^{\rm min}_{ |p|}=\frac{\nu}{2}\,,\qquad \Delta^{\rm min}_{ 1}=\frac{1}{2}\left[\f{\nu}{p^{2}}\!\!+  (1-\frac{\xi }{p})\right].
\label{scaling1}
\ee 
Note that among these two, the $|p|$-agglomerate is always the
most relevant since $\Delta^{\rm min}_{|p|}<\Delta^{\rm min}_{1}$ with the only exception for $\nu=2/3$ in the LW model
($\xi=-1$), where both have equal scaling \cite{Wen92,
  Kane95}.  At higher energies $\omega_{\rm n}\ll E\ll \omega_{\rm c}$
the neutral mode saturates and does not contribute to the scaling $\Delta_{m}$, which consequently depends on the charged mode only with a
value $\Delta^{\rm eff}_m={\nu m^2}/{2 p^2}$.  Here, the single-qp
($m=1$) always dominate. This implies the possibility of a crossover regime from low
energies (relevance of $|p|$-agglomerates) to higher energies (relevance
of single-qp).  In the presence of interactions, Eq.(\ref{scaling})
shows the relevance of the $|p|$-agglomerate at low energies if
$g_{\rm{n}}/g_{\rm{c}}>\nu (1+\xi/p)$, otherwise the single-qp will always
dominate.

\section{Transport properties} 
Tunneling of a bunched $m$-excitations through the QPC located  at $x=0$ is 
described by $H^{(m)}_{\rm{T}}=\textbf{t}_{m}{\Psi^{(m)}_{\rm{R}}}^{\dagger}(0)\Psi^{(m)}_{\rm{L}}(0)+{\rm h.c.}$ with amplitude $\textbf{t}_{m}$. The indices $\rm R$ and $\rm L$ represent the right and left edge of the 
Hall bar. We will consider only the relevant excitations with $m= 1 $ (single-qp) or $m= |p|$ ($|p|$-agglomerate).
In the incoherent sequential regime and at lowest order in $H^{(m)}_{\rm{T}}$ 
higher current cumulants $\langle I^{(m)}_{\rm{B}}\rangle_{k}$ ($k$-th order cumulant) are expressed in terms of the backscattering current $I^{(m)}_{\rm{B}}$ 
\be
\langle I^{(m)}_{\rm{B}}\rangle_{k} =\left\{
\ba
(m e^{*})^{k-1}\coth\left(E_{m}/2T\right)I^{(m)}_{\rm{B}}\\
(m e^{*})^{k-1}I^{(m)}_{\rm{B}}
\ea
\right.
\ba
k \,\,\,{\rm{even}}\\
k \,\,\,{\rm{odd}}
\ea
\label{cumulant}
\ee
since the statistics is bidirectional Poissonian\cite{Levitov04}.
The current is proportional to the tunneling rate $\Gamma^{(m)}(E)$ as $I^{(m)}_{\rm{B}}=m e^{*}(1-e^{-E_{m}/T})\Gamma^{(m)}(E_{m})$ with 
\be
\Gamma^{(m)}(E_{m})=
\gamma_m^2\!\int^{+\infty}_{-\infty}\!\!\!\!\!\! d t' e^{-iE_m t'} e^{2\alpha^{2}_{m} 
\mathcal{D}^{>}_{\rm{c}}(t')}e^{2\beta^{2}_{m} \mathcal{D}^{>}_{\rm{n}}(t')}\,.
\label{Rate}
\ee 
Here, $E_{m}=m e^{*}V$, with $V$ the QPC bias voltage and
$\gamma_m^2=|\textbf{t}_{m}|^{2}/(4\pi^{2} a^{2})$. The charge
coefficient is $\alpha_{m}=m/|p|$ while the neutral one is given by
the minimal value with $q=0$ in Eq.(\ref{qp_operator}). For the
single-qp it is $\beta_1^2=(1-\xi/p)$, while for the
$|p|$-agglomerate it is $\beta_{|p|}=0$.  The correlation functions
\cite{Braggio01,Ferraro08} of charged and neutral modes are 
\be
\mathcal{D}^{>}_{r}(t)=g_{r}\nu_{r}\ln{\left[\frac{|\mathbf{\Gamma}\left(1+T/\omega_{r}-iTt\right)|^{2}}{\bold{\Gamma}^{2}\left(1+T/
        \omega_{r}\right) \left(1-i\omega_{r}t\right)}\right]},
\label{correlation}
\ee
with $r=\rm{c},\rm{n}$ and $\bold{\Gamma}(x)$ the Euler Gamma function. The rate  is obtained by numerically evaluating (\ref{Rate})
apart at zero temperature where analytical  results are available~\cite{Ferraro08}.

At lowest order, tunneling processes of different excitations are
independent. The contributions of different excitations
are then simply summed. In our case, the total $k$-th order cumulant will be given by the sum of the most relevant processes
$\langle I_{\rm{B}}\rangle_{k}=\langle
I^{(1)}_{\rm{B}}\rangle_{k}+\langle I^{(p)}_{\rm{B}}\rangle_{k}$.  The
trasmission of the QPC is then expressed in terms of the total
backscattering current \be t=1-I_{\rm{B}}/I_{0}\,,\qquad {\rm
  with}\qquad I_{0}=(\nu e^{2}/2\pi)V\,,
\label{transmission}
\ee where, for simplicity, we denoted $I_{\rm{B}}\equiv\langle
I_{\rm{B}}\rangle_{1}$. Among higher cumulants, backscattering current
noise is an essential quantity in order to extract information on
charge excitations. It consists of the excess backscattered noise
$S_{\rm{exc}}$, due to finite current, and the thermal Johnson-Nyquist
noise 
\be \langle I_{\rm{B}}\rangle_{2}=S_{\rm{exc}}+2T
G_{\rm{B}}(T)\,, \ee 
with $G_{\rm{B}}$ the total backscattering conductance \cite{Note2}.  
Note that, at lowest order in tunneling, the backscattered
excess noise coincides with the transmitted excess noise which is
usually measured in experiments\cite{Ponomarenko99,Dolcini05}. For this reason,
treating the high transmission regime, we will analyze $S_{\rm{exc}}$ and we will compare it
with experiments.  

Often in experiments it is
introduced the effective charge, $e_{\rm{eff}}(T)$, defined as the
\emph{single} carrier that better fits the excess noise at a given
temperature $T$ \cite{Chung03,Ofek09} \be
\label{Sexceff}
S_{\rm{exc}}=e_{\rm{eff}}(T)\coth\left[\f{e_{\rm{eff}}(T)V}{2T}\right]I_{\rm{B}}(V,T)-2T
G_{\rm{B}}(T).  
\ee 
One has to be aware that this quantity has a clear meaning of
real tunneling charge when is guaranteed the presence of a single
dominant carrier, otherwise it represents a weighted average of different carriers. Its value strongly depends on the voltage range
considered. 

In the shot noise regime $e^*V\gg T$ it is 
\be {e}^{\rm sh}_{\rm{eff}}=e^*
\frac{I^{(1)}_{\rm{B}} +|p| I^{(p)}_{\rm{B}}}{ I_{\rm{B}}}\,.  \ee 
In
the opposite regime, $e^*V<T$, often considered in experiments, it can be deduced from the behavior of (\ref{Sexceff})
in the limit $V\to 0$
\be
{e}^{\rm th}_{\rm{eff}}(T)=\left[\f{3T}{G^{(\rm{tot})}_{\rm{B}}} \left(\f{d^{2}S_{\rm{exc}}}
{dV^{2}}-\f{2}{3}T\f{d^{3} I^{}_{\rm{B}}}{dV^{3}} \right)\right]^{\f{1}{2}}_{V\to 0}.
\label{charge}
\ee
IUsing the relation (\ref{cumulant}) this effective charge can be equivalently expressed  in terms of the third order cumulant \be
{e}^{\rm th}_{\rm{eff}}(T) =e \left[\f{\langle
    I_{\rm{B}} \rangle_{3}}{(e^{2}
    I_{\rm{B}})}\right]^{\f{1}{2}}_{V\rightarrow 0}.
\label{eff_skew}
\ee This corresponds to the square root of the normalized
skewness at zero voltage~\cite{Ferraro09b} and it can be interpreted as the definition of the 
effective charge in the thermal regime. This quantity can be compared with the effective charge 
measured in the experiments as a function of temperature.
\section{Results} 
In this part we will focus on the comparison with available
experimental data for $\nu=2/5$ ($p=2$) and $\nu=2/3$ ($p=-2$).
Parameters are chosen in order to guarantee a crossover between the
$|p|$-agglomerate at low energies and the single-qp at higher
energies. Figures and fitting will be presented for the LW model
$\xi={\rm{sgn}}(p)$, which corresponds to a
counter-propagating (co-propagating) neutral mode for $\nu=2/3$
($\nu=2/5$).  The opposite case of $\xi=-{\rm{sgn}}(p)$ (GFL model) is
straighforwardly obtained using the mapping
(\ref{eq:mapping}).
 
At low temperature $T\ll e^* V$ (shot noise regime) the total current
and the excess noise show similar power law behavior
$I_{\rm{B}}\propto V^{\eta-1}$, $S_{\rm{exc}}\propto V^{\eta-1}$ with
scaling exponent $\eta$  depending on the voltage regimes (see below)
\be
\hskip-0.1cm\eta_1\!=\!{2 g_{\rm{c}}\nu};\;\; \eta_2\!=\!{2g_{\rm{c}}\f{\nu}{p^{2}}+2 g_{\rm{n}}\left(\!1\!-\!\f{\xi}{p}\!\right)};\;\; \eta_3\!=\!{2g_{\rm{c}} \f{\nu}{p^{2}}}\,.
\label{eta}
\ee For $V\ll V^*$, $|p|$-agglomerates dominate with $\eta=\eta_1$. At
higher voltages, $V^*\ll V\ll \omega_{\rm n}/e^*$ single-qps become more
relevant and neutral modes contribute to the dynamics with
$\eta=\eta_2$ . At even higher bias $V\gg \omega_{\rm n}/e^*$ the neutral
modes saturate giving $\eta=\eta_3$. The crossover voltage $V^*$ is defined as
the bias at which the two current contributions are equal
$I_{\rm{B}}^{(1)}(V^*)=I_{\rm{B}}^{(p)}(V^*)$. The explicit value depends on 
intrinsic parameters such as the ratio of the tunneling amplitudes $\gamma_2/\gamma_1$ \cite{Ferraro09b}.

At higher temperature $T \gg e^{*}V$ (thermal regime) the current is linear in
voltage with a temperature dependent total backscattering conductance
$G_{\rm{B}}(T)\propto T^{\eta-2}$. The scaling exponent varies as
function of temperature, with $\eta=\eta_1$ for $T\ll T^*$,
$\eta=\eta_2$ for $T^*\ll T\ll\omega_{\rm n}$, and $\eta=\eta_3$ for
$T\gg\omega_{\rm n}$. The crossover temperature $T^*$ separates the
region of relevance between the $|p|$-agglomerate and the single-qp in
the linear conductance. Its value depends explicitly on the model
parameters such as interaction renormalizations and amplitude ratio
$\gamma_{2}/\gamma_{1}$. It corresponds to the value where
$G_{\rm{B}}^{(p)}(T^*)=G_{\rm{B}}^{(1)}(T^*)$.  In the same regime the excess
noise is quadratic in the bias $S_{\rm{exc}}\propto V^{2}$.

\begin{figure}
	\centering
		\includegraphics[width=0.45\textwidth]{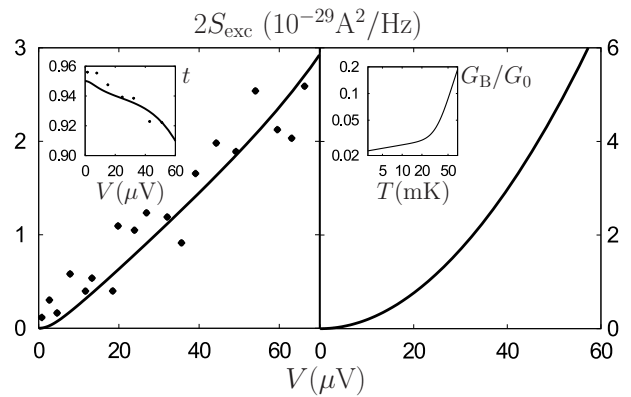}
		\caption{(a) Excess noise at $\nu=2/3$ (in unit of
                  $10^{-29}$ ${\rm{A}}^{2}/{\rm{Hz}}$) as a function
                  of $V$ for $T=10$ mK
                  (corresponding voltage $V=T/e^*=2.6$ $\mu V$).  Inset:
                  transmission $t$ as given in Eq. (\ref{transmission}) as a function of $V$ with
                  $t(V=0)=0.95$. Diamonds represent the experimental
                  data taken from Ref.~\onlinecite{Ofek09}, with
                  courtesy of Moty Heiblum.  (b) Same as in (a) but
                  at $T=80$ mK.  Inset: log-log
                  plot of the total linear backscattering conductance (in
                  unit of $G_{0}=e^{2}/2 \pi$) as a function of temperature. Other parameters: $g_{\rm{c}}=1.6$,
                  $g_{\rm{n}}=8.1$, $\omega_{\rm{c}}=5$ K,
                  $\omega_{\rm{n}}=200$ mK,
                  $\gamma_{2}/\gamma_{1}=0.20$, $\gamma_1^2 /
                  \omega_c^2=1.1\cdot 10^{-1}$.}
\label{Fig1}
\end{figure}

Fig. \ref{Fig1}a shows the excess noise and the QPC transmission as a
function of the external voltage for $\nu=2/3$ at extremely low
temperature $T=10$ mK.  The parameters are chosen in order to
fit the experimental data (black diamonds)~\cite{Ofek09}.  The
voltages considered are mainly in the shot noise regime, $e^* V>T$.  The
excess noise shows an almost linear behavior until very small voltages with a single
power law. We then select the $m=|p|=2$ contribution, which is the relevant at low energies, 
with  $S_{\rm{exc}}\propto V^{\eta_1-1}$ and ${e}^{\rm
  sh}_{\rm{eff}}=2e/3$.  The fit of the experimental data fixes the
interaction to $g_{\rm c}=1.6$ (cf. Eq.(\ref{eta})). This value is
also used to plot the transmission in (\ref{transmission}) as shown in the
inset. A good agreement with the data is visible. 
Note that having considered the contribution of the $|p|$-agglomerate it 
fixes a lower bound to the crossover voltage that has to be higher 
than the voltage's window considered $V^*> 70 \mu$V.  In order to
obtain informations on the single-qp one should investigate higher
voltage or temperature regimes.  In Fig.\ref{Fig1}b, main panel, we
show the expected higher temperature noise for $T=80$ mK. For $V \ll
T/e^*\approx 21 \mu$V the parabolic behavior of the thermal excess
noise is visible. In the same regime the current is linear in voltage
with a temperature dependent conductance (see inset). Here, the
temperature range is chosen in order to show the first two scaling
regimes: from $\eta_1$ ($|p|$-agglomerate) to $\eta_2$ (single-qp),
indeed we have $T^*=42$ mK. Note that the noise behavior in the main figure is at $T>T^*$,
where single-qp tunneling processes dominate. This is confirmed by the value of 
effective charge given by ${e}^{\rm {th}}_{\rm{eff}}=e/3$.
\newline
The above results demonstrates that the value of the effective
tunneling charge crucially depends on the external parameters such as
temperature and voltage.  

\begin{figure}
	\centering
	\includegraphics[width=0.45\textwidth]{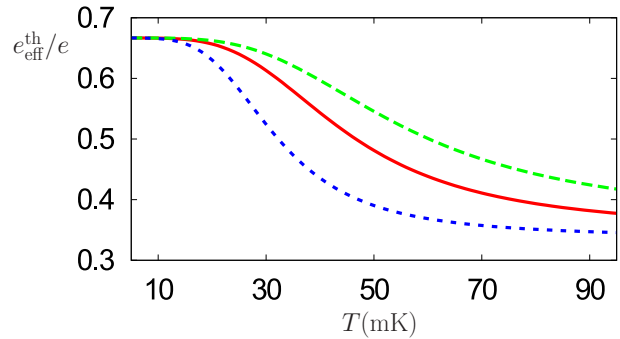}
	 \caption {Effective charge, in unit of the electron charge $e$, as a function of 
temperature, for $\nu=2/3$  and  different values of the ratio  $\gamma_{2}/\gamma_{1}=0.1$ (blue, short-dashed), $0.2$ (red, straigth), $0.35$ (green, long-dashed). The corresponding crossover temperatures are 
$T^*=32$ mK, $42$ mK, $60$ mK respectively. The other parameters are as in Fig. \ref{Fig1}. }
	\label{Fig2}
\end{figure}

This point can be further analyzed by
considering the temperature dependence of the effective charge at low
voltages, $e^*V<T$.
Fig. \ref{Fig2} shows ${e}^{\rm th}_{\rm{eff}}$, evaluated using the
expression (\ref{eff_skew}), for different values of the tunneling amplitude ratio $\gamma_{2}/\gamma_{1}$  between a bunch of two qps $(\gamma_{2})$  and a single qp $(\gamma_{1})$.  At low
temperatures, the effective charge corresponds to the
$|p|=2$ agglomerate with ${e}^{\rm th}_{\rm{eff}}= \nu e$, while,
increasing temperature, it reaches the single-qp value ${e}^{\rm
  th}_{\rm{eff}}=\nu e/|p|$. The crossover region between
the two regimes is driven by $T^*$ which increases increasing the ratio of  $\gamma_{2}/\gamma_{1}$.

We conclude the comparison with experiments by considering the effective charge for filling factor 
$\nu=2/5$ where experimental data are available. This case was discussed in Ref.~\onlinecite{Ferraro08} where model parameters were fixed by
fitting the temperature dependence of the linear conductance. Here we
focus on the temperature behavior of the effective charge.
\begin{figure}
	\centering
	\includegraphics[width=0.45\textwidth]{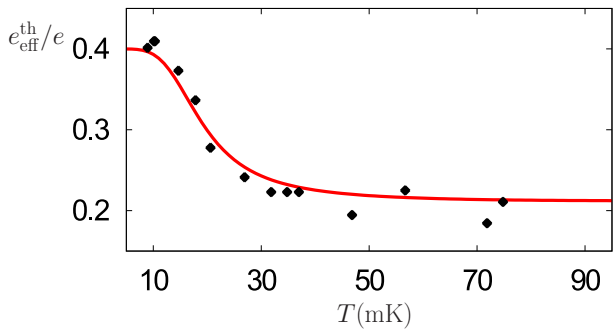}
	 \caption {Effective charge, in unit of the electron charge $e$, as a function of 
temperature, for $\nu=2/5$. Diamonds represent the experimental data taken from Ref.~\onlinecite{Chung03}, with courtesy of Moty Heiblum.  Parameters: $g_{\rm{c}}=3$, $g_{\rm{n}}=12$, $\omega_{\rm{c}}=5$ K, $\omega_{\rm{n}}=50$
mK, $\gamma_{2}/\gamma_{1}=0.65$, with  $T^*=18$ mK.}
	\label{Fig3}
\end{figure}
Fig. \ref{Fig3} shows the evolution of ${e}^{\rm th}_{\rm{eff}}$ as a
function of temperature. The agreement with the corresponding
quantity measured in Ref.~\onlinecite{Chung03} (black diamonds)
is very good and reinforces the crossover scenario of tunneling from single-qps to agglomerates at sufficiently low 
temperature. Note that
for the above fit we used the parameters fixed in Ref.~\onlinecite{Ferraro08} for the linear
conductance. They are however here expressed for the LW model with co-propagating neutral and charged modes\cite{Ferraro08}.

\section{Conclusion}
We proposed a minimal hierarchical model which fully explains recent
experimental observations on excess noise at low
temperatures and weak backscattering. The meaning of the effective
charge and its temperature dependence was analyzed in comparison
with the available experimental data. A quantitative analysis of the
dependence of noise and effective charge on external parameters was performed. Evidence of
neutral modes propagating with finite velocity and quantitative value of the corresponding bandwidth were extracted.  

Our results show that the increasing of the effective charges, observed in experiments at extremely low temperatures
for the Jain sequence, can be well explained in 
terms of the dominance of the $|p|$-agglomerates over the single-qp contribution. Only at sufficiently 
high energies the single-qp dominance is again recovered. We expect that the described crossover could 
be also relevant for other filling factors, outside of the Jain sequence, where anomalous 
increasing of the effective charges is also observed \cite{dolev10}.   

As a final remark we note that within the analyzed geometry with a point-like scatterer 
we cannot shed light on the propagation direction of the neutral modes, but only on their presence. 
 The fit of the experiments were done using the value $\xi={\rm{sgn}}(p)$ (LW model),
which corresponds to a counter-propagating neutral mode for $\nu=2/3$ in accordance with recent observations\cite{Aveek10}.
However, one could have fit as well the data in the other case with
$\xi=-{\rm{sgn}}(p)$ (GFL model) with a co-propagating neutral mode
for $\nu=2/3$, simply changing the interaction parameters (cf. Eq.(\ref{eq:mapping})).  Anyway, to have
information on the direction of propagation one should consider more complicated geometries such as the four terminal steup recently addressed in experiments \cite{Aveek10}.

\vskip0.7cm

\section*{ACKNOWLEDGEMENT}

We thank M. Heiblum, M. Dolev, N. Ofek and A. Bid for valuable discussions on the 
experiments and A. Cappelli, G. Viola and M. Carrega for useful discussions.
Financial support of the EU-FP7 via ITN-2008-234970 NANOCTM
is gratefully acknowledged.


\begin{thebibliography}{10}
\bibitem{DasSarma97} S. Das Sarma, A. Pinczuk \emph{Perspective in Quantum Hall Effects: Novel Quantum Liquid in
Low-Dimensional Semiconductor Structures} (Wiley, New York 1997).
\bibitem{Laughlin83} R. B. Laughlin, Phys. Rev. Lett \textbf{50}, 1395 (1983).
\bibitem{Wen90}  X. G. Wen, Phys. Rev. Lett. \textbf{64}, 2206 (1990).
\bibitem{Wen91} X. G. Wen, Phys. Rev. B \textbf{43}, 11025 (1991).
\bibitem{Wen95} X. G. Wen, Adv. Phys. \textbf{44}, 405 (1995).
\bibitem{dePicciotto97} R. de Picciotto, M. Reznikov, M. Heiblum, V. Umansky, G. Bunin, and  D. Mahalu, {Nature} \textbf{389}, 162 (1997).
\bibitem{Seminadayar97} L. Saminadayar, D. C.  Glattli, Y. Jin, and B. Etienne, {Phys. Rev. Lett.} \textbf{79}, 2526 (1997).
\bibitem{Chung03} Y.  C. Chung, M. Heiblum, and V. Umansky, {Phys. Rev. Lett} \textbf{91}, 216804 (2003).
\bibitem{Ofek09} A. Bid, N. Ofek, M. Heiblum, V. Umansky, and D. Mahalu, {Phys. Rev. Lett.} \textbf{103}, 236802 (2009).
\bibitem{dolev10} M. Dolev,  Y. Gross, Y. C. Chung,  M. Heiblum, V. Umansky, and D. Mahalu, {Phys. Rev. B} {\bf{81}}, 161303(R) (2010).
\bibitem{Roddaro04} S. Roddaro, V. Pellegrini, F. Beltram, G. Biasiol, and L. Sorba, Phys. Rev. Lett. \textbf{93}, 046801 (2004).
\bibitem{Fendley95} P. Fendley, A. W. W. Ludwig, and H. Saleur,  Phys. Rev. Lett. {\bf 75}, 2196 (1995).
\bibitem{Rosenow02} B. Rosenow and  B. I. Halperin, Phys. Rev. Lett. \textbf{88}, 096404 (2002).
\bibitem{Papa04} E. Papa and  A. H. MacDonald, Phys. Rev. Lett. \textbf{93}, 126801 (2004).
\bibitem{Mandal02} S. S. Mandal and  J. K. Jain, Phys. Rev. Lett. \textbf{89}, 096801 (2002).
\bibitem{Yang03} K. Yang,  Phys. Rev. Lett.  \textbf{91}, 036802 (2003).
\bibitem{Aleiner94} I. L. Aleiner and L. I. Glazman, Phys. Rev. Lett. \textbf{72}, 2935 (1994).
\bibitem{Wen92} X. G. Wen and  A. Zee, {Phys. Rev. B} \textbf{46}, 2290 (1992).
\bibitem{Kane94} C. L. Kane,  M. P. A. Fisher, and J. Polchinski, Phys. Rev. Lett. \textbf{72}, 4129 (1994).
\bibitem{Kane95} C. L. Kane and  M. P. A. Fisher, Phys. Rev. B \textbf{51}, 13449 (1995).
\bibitem{Lopez99} A. Lopez and E. Fradkin, Phys. Rev. B {\bf{59}}, 15323 (1999).
\bibitem{Lopez01} A. Lopez and E. Fradkin, Phys. Rev B \textbf{63}, 085306 (2001).
\bibitem{Ferraro08} D. Ferraro, A. Braggio, M. Merlo, N. Magnoli, and M. Sassetti,  Phys. Rev. Lett. {\bf{101}}, 166805 (2008).
\bibitem{Levin07} M. Levin, B. I. Halperin, and B. Rosenow,  Phys. Rev. Lett. {\bf{99}}, 236806 (2007).
\bibitem{Feldman08} D. E. Feldman and F. Li,  Phys. Rev. B. {\bf{78}}, 161304 (2008). 
\bibitem{Overbosch09} B. J. Overbosch and  C. Chamon, {Phys. Rev. B} \textbf{80}, 035319 (2009).
\bibitem{Cappelli09} A. Cappelli, L. S. Georgiev, and G. R. Zemba, J. Phys. A: Math. Theor. \textbf{42}, 222001 (2009).
\bibitem{Cappelli10} A. Cappelli, G. Viola, and G. R. Zemba, Ann. Phys. \textbf{325}, 465 (2010).
\bibitem{Ferraro09b} D. Ferraro, A. Braggio, N. Magnoli, and M. Sassetti, {New J. Phys.} \textbf{12}, 013012 (2010).
\bibitem{Yang09} Z. X. Hu, E. H. Rezayi, X. Wan, and K. Yang, Phys. Rev. B {\bf{80}}, 235330 (2009). 
\bibitem{Granger09} G. Granger, J. P. Eisenstein, and J. L. Reno,  Phys. Rev. Lett. {\bf{102}}, 086803 (2009).
\bibitem{Aveek10} B. Aveek, N. Ofek, H. Inoue, M. Heiblum, C. L. Kane, V. Umansky, and D. Mahalu, cond-mat/1005.5724 (unpublished).
\bibitem{Reulet03} B. Reulet, J. Senzier, and D. E. Prober, Phys. Rev. Lett. {\bf 91}, 196601 (2003).
\bibitem{Lindell04} R. K. Lindell, J. Delahaye, M. A. Sillanp\"a\"a, T. T. Heikkil\"a, E. B. Sonin, and P. J. Hakonen,
Phys. Rev. Lett. {\bf 93}, 197002 (2004).
\bibitem{Bomze05} Yu. Bomze, G. Gershon, D. Shovkun, L. S. Levitov, and M. Reznikov, Phys. Rev. Lett. {\bf 95}, 176601 (2005).
\bibitem{Huard07} B. Huard, H. Pothier, N. O. Birge, D. Esteve, X. Waintal, and J. Ankerhold, Ann. Phys. {\bf 16}, 736 (2007).
\bibitem{Timofeev07}  A. V. Timofeev, M. Meschke, J. T. Peltonen, T. T. Heikkil\"a, and J. P. Pekola, Phys. Rev. Lett. {\bf 98}, 207001 (2007).
\bibitem{Gershon08}  G. Gershon, Yu. Bomze, E. V. Sukhorukov, and M. Reznikov, Phys. Rev. Lett. {\bf 101}, 016803 (2008).
\bibitem{Levkinskyi08} I. P. Levkivskyi and  E. V. Sukhorukov,  {Phys. Rev. B} {\bf{78}}, 045322 (2008).
\bibitem{Levkinskyi09} I. P. Levkivskyi, A. Boyarsky, J. Frohlich, and E. V. Sukhorukov, {Phys. Rev. B} {\bf{80}}, 045319 (2009).
\bibitem{Lee98} D. H.  Lee and X. G. Wen, cond-mat/9809160 (unpublished).  
\bibitem{Chamon07} C. Chamon, E. Fradkin, and A. Lopez, {Phys. Rev. Lett.} \textbf{98}, 176801 (2007).
\bibitem{Note1} For $m<0$ one describes quasi-hole excitations not considered here.
\bibitem{Su86} W. P. Su,  {Phys. Rev. B} \textbf{34}, 1031 (1986).
\bibitem{Froehlich97} J. Fr\"ohlich, U. M. Studer, and E. Thiran,  J. Stat. Phys. \textbf{86}, 821 (1997).
\bibitem{Ino98} K. Ino, {Phys. Rev. Lett.} \textbf{81}, 5908 (1998).
\bibitem{Guyon02} R. Guyon, P. Devillard, T. Martin, and I. Safi,  Phys. Rev. B {\bf{65}}, 153304 (2002).
\bibitem{Martin05} T. Martin, \emph{Les Houches Session LXXXI} ed. H. Bouchiat \emph{et al.} (Elsevier, Amsterdam, 2005).
\bibitem{Kane92} C. L. Kane and  M. P. A. Fisher, {Phys. Rev. Lett.} \textbf{68}, 1220 (1992).
\bibitem{Levitov04} L. S. Levitov and  M. Reznikov, {Phys. Rev. B} \textbf{70}, 115305 (2004).
\bibitem{Braggio01} A. Braggio, M. Sassetti, and B. Kramer, Phys. Rev. Lett. \textbf{87}, 146802 (2001).
\bibitem{Note2} Note that our definition of noise differs from the one usually considered in experimental works (see e.g. \onlinecite{Chung03, Ofek09})  by a factor $2$ that has been properly taken into account in the analysis of the data. This definition is useful in order to avoid multiplicative factors in the expression of the current cumulants.
\bibitem{Ponomarenko99} V. V. Ponomarenko and N. Nagaosa, Phys. Rev. B {\bf 60}, 16865 (1999).
\bibitem{Dolcini05} F. Dolcini, B. Trauzettel, I. Safi, and H. Grabert, Phys. Rev. B \textbf{71}, 165309 (2005).
%\bibitem{Griffiths00} T. G. Griffiths, E. Comforti, M. Heiblum, A. Stern, and V. Umansky, Phys. Rev. Lett. {\bf 85} , 3918 (2000). 
\end{thebibliography}
\end{document}